\documentclass{PoS}

\usepackage{graphicx}
\usepackage{amsmath,amssymb,amsfonts}
\usepackage{psfrag}
\usepackage{bm}
\usepackage{xcolor}
\let\normalcolor\relax

\newcommand{\vc}[1]{{\bm{#1}}}
\newcommand{\ev}[1]{\left\langle #1 \right\rangle}
\newcommand{\nN}[1]{\left\langle #1 \right\rangle}
\newcommand{\abs}[1]{\left| #1 \right|}

\newcommand{\Seff}{S_{\text{eff}}}
\newcommand{\Sw}{S_{\text{W}}}
\newcommand{\YM}{\text{YM}}

\newcommand{\coco}{\mathrm{c.c.}}
\newcommand{\bbZ}{\mathbb{Z}}
\newcommand{\trP}{\mathcal{P}}
\newcommand{\Ns}{N_\text{s}}
\newcommand{\Nt}{N_\text{t}}
\newcommand{\ord}{\mathcal{O}}
\newcommand{\vcx}{\vc{x}}
\newcommand{\vcy}{\vc{y}}
\newcommand{\rep}{\mathcal{R}}
\newcommand{\Ed}{E_{\text{D}}}

\DeclareMathOperator{\tr}{tr}
\renewcommand{\Re}{\mathrm{Re\,}}
\renewcommand{\Im}{\mathrm{Im\,}}
\newcommand{\Prot}{P_\text{rot}}

\definecolor{doran}{rgb}{1,0.2,0}
\definecolor{viol}{rgb}{0.5,0,1}

\newcommand{\myindent}{\noindent}

\title{Inverse Monte-Carlo and Demon Methods for Effective
Polyakov Loop Models of SU(N)-YM}

\ShortTitle{IMC and Demon Methods for Effective
Polyakov Loop Models of $SU(N)$-YM}

\author{\speaker{Christian Wozar}, Tobias~K\"astner, Bj\"orn~H.~Wellegehausen, Andreas~Wipf\\
        Theoretisch-Physikalisches Institut,
Friedrich-Schiller-Universit{\"a}t Jena, Max-Wien-Platz 1, 07743
Jena, Germany\\
        E-mail: \email{christian.wozar@uni-jena.de}}

\author{Thomas~Heinzl\\
       School of Mathematics and Statistics, University of
Plymouth, Drake Circus, Plymouth, PL4~8AA, United Kingdom}

\abstract{We study effective Polyakov loop models for $SU(N)$
Yang-Mills theories at finite temperature. In particular effective models for
$SU(3)$ YM with an additional adjoint Polyakov loop potential are considered.
The rich phase structure including a center and anti-center directed phase is
reproduced with an effective model utilizing the inverse Monte-Carlo method. 
The demon method as a possibility to obtain the effective models' couplings is
compared to the method of Schwinger-Dyson equations. Thermalization effects of
microcanonical and canonical demon method are analyzed. Finally the elaborate 
canonical demon method is applied to the finite temperature $SU(4)$ YM phase
transition.}

\FullConference{The XXVI International Symposium on Lattice Field Theory\\
                 July 14-19 2008\\
                 Williamsburg, Virginia, USA}

\begin{document}

\section{Introduction}
\myindent
\setlength{\baselineskip}{0.85\baselineskip}%
The Svetitsky-Yaffe conjecture
\cite{Svetitsky:1982gs} states that the
Yang-Mills finite temperature transition in dimension $d+1$ is
described by an effective spin model in $d$ dimensions with
short range interactions. This relationship is analyzed for a $SU(3)$ YM theory
with adjoint Polyakov loop potential and the $SU(4)$ YM theory in terms of
inverse Monte-Carlo (IMC) methods. Ways to perform IMC are provided by demon
methods \cite{Creutz:1984fj,Gocksch:1984ih}. Specific thermalization effects of
these must be discussed to obtain reliable results. Taking these effects into
account we compare the corresponding
results to a Schwinger-Dyson approach to IMC \cite{Falcioni:1985bh}.

\section{Effective models for Yang-Mills theories}
\myindent
We start with the well-known lattice Wilson action
\begin{equation}
 \Sw = \beta \sum_{\square} \left( 1-  \frac{1}{N_\text{C}} \Re \, \tr
 \, U_{\square} \right),\quad \beta = \frac{6}{a^4g^2}
\end{equation}
and perform a \emph{strong coupling expansion} (for small
$\beta$). Since the resulting `operators' (Polyakov loop
monomials) are \emph{dimensionless} there is no natural
ordering scheme. We therefore use a truncation scheme based on the ordering by
powers of $\beta$ which are closely related to the dimension of the corresponding group
representations and ordering by the distance across which the Polyakov
      loops are coupled.
In compact form the strong coupling expansion is given by
\begin{equation}
  \Seff = \sum_r \sum_{\rep_1 \ldots  \rep_r} \sum_{\ell_1 \ldots \ell_r}
  c_{\rep_1 \ldots \rep_r}^{\ell_1 \ldots  \ell_r}(\beta)  \prod_{i=1}^r
  S_{\rep_i, \ell_i} = \sum_i \lambda_i S_i
\end{equation}
with the basic building blocks
\begin{equation}
  S_{\rep, \ell} \equiv \chi_\rep (\trP_\vcx) \chi_\rep^* (\trP_{\vcy}) + \coco
  , \quad \ell \equiv \nN{\vcx\vcy} .
\end{equation}
Here $r$ counts the number of link operators contributing at
each order. The coefficients $c_{\rep_1 \ldots \rep_r}^{\ell_1
\ldots \ell_r}$ are the couplings between the operators
$S_{\rep_i,\ell_i}$ sitting at nearest-neighbor (NN) links
$\ell_i \equiv \nN{\vcx_i,\vcy_i}$ in representation $\rep_i$.
The effective action hence describes a \emph{network of link
operators} that are collected into (possibly disconnected)
`polymers' contributing with `weight' $c_{\rep_1 \ldots
\rep_r}^{\ell_1 \ldots \ell_r}$. One expects the `weights' or
couplings to decrease as the dimensions of the involved
representations and inter-link distances increase. In a strong
coupling (small $\beta$) expansion truncated at $\ord(\beta^{k
\Nt})$ one has $r \le k$ and the additional restriction
$|\rep_1| + \cdots + |\rep_r| < k$ with $|\rep| \equiv \sum_i p_i $
for a given representation $\rep$ of $SU(N)$ with Dynkin labels
$[p_1,\ldots,p_N]$.

\subsection*{Effective models for $\mathbf{SU(3)}$ Yang-Mills}
\myindent
To lowest order $\ord(\beta^{\Nt})$ one finds the universal effective action
\begin{equation}
\Seff = c_{10} \sum_{\nN{\vcx\vcy}} S_{10, \nN{\vcx\vcy}}
\equiv \lambda_1
\sum_{\nN{\vcx\vcy}} (\trP_\vcx  \trP_\vcy^*  +  \trP_\vcx^* \trP_\vcy ).
\end{equation}
Our truncated model to order $\ord(\beta^{2\Nt})$ and with \emph{nearest
neighbor interactions} reads as 
\begin{equation}
\label{eq:su3model}
\begin{aligned}
\Seff &= \lambda_1
\sum_{\nN{\vcx\vcy}}(\chi_{10}(\trP_\vcx)\chi_{01}(\trP_\vcy)+\coco)
 + \lambda_2 \sum_{\nN{\vcx\vcy}}(\chi_{20}(\trP_\vcx)\chi_{02}(\trP_\vcy)+\coco)\\
 &\quad + \lambda_3
\sum_{\nN{\vcx\vcy}}(\chi_{10}(\trP_\vcx)\chi_{01}(\trP_\vcy)+\coco)^2 .
\end{aligned}
\end{equation}
For a discussion of effective $SU(3)$ Polyakov loop models see
\cite{Wozar:2006fi,Wipf:2006wj}.

\section{Inverse Monte-Carlo -- the basics}
\myindent
The inverse Monte-Carlo (IMC) method \cite{Falcioni:1985bh} allows to determine 
(effective) actions from given configurations. In our case, these are Polyakov
loops obtained from gauge configurations generated with the
Wilson action. Via IMC we determine the couplings of
truncated effective actions which (ideally) would give rise to
the same distribution of Polyakov loop configurations.

The IMC procedure is based on an \emph{ansatz} for the
effective action of the type $\Seff = \sum_i \lambda_i S_i$.
Translational invariance of the reduced Haar measure leads to
Schwinger-Dyson (SD) equations \cite{Uhlmann:2006ze}. They constitute an
\emph{overdetermined} linear system for the effective couplings
$\lambda_i$ which may be solved by least-square methods (see
\cite{Heinzl:2005xv,Wozar:2007tz,Wozar:2007eu}). A second way to determine the couplings
$\lambda_i$ is the \emph{demon method} which was successfully applied to
$SU(2)$ YM in \cite{Velytsky:2008bh}.

\section{Microcanonical demon method}
\myindent
Based on the large volume relation between microcanonical and canonical ensemble
in statistical physics additional degrees of freedom (\emph{``demons''}) are
used to simulate the effective theory together with the demons at a fixed total
energy/action \cite{Creutz:1983ra}. The demon is used as a ``thermometer'' to \emph{measure the
coupling} of the corresponding part of the effective action.

To simulate a microcanonical system with action $S[\trP]=\sum_i \lambda_i S_i[\trP]$ we
transform the canonical measure to the microcanonical one,
\begin{equation}
\rho[\trP,\Ed] \propto \exp\Bigl(-\sum_i \lambda_i (S_i[\trP]+\Ed^i)\Bigr)
\quad \longrightarrow \quad
\delta[S_i[\trP]+\Ed^i-E_\text{total}^i] .
\end{equation}

Each demons' energy $\Ed^i$ is
distributed according to $\rho(\Ed^i) \propto \exp(-\lambda_i \Ed^i)$, 
$\lambda_i$ depending on $\ev{S_i}$.
The constraint $\Ed^i \in [-E_0^i, E_0^i]$, $E_0^i<\infty$
leads to an invertible relation $\ev{\Ed^i} = f_i(\lambda_i)$. 
Couplings are obtained via the demon method in the following way:
\begin{enumerate}
  \setlength{\itemsep}{0pt}
  \item Simulate the microscopic (full YM) system without additional demons.
  \item Reduce the system for a chosen configuration to
  a Polyakov loop configuration. 
  \item Perform a microcanonical simulation of the reduced Polyakov loop system with
  coupled demons. As discussed below the thermalization procedure should be
  handled cautious.
  \item The mean energy of the demons is directly related to the couplings
  of the effective theory.
\end{enumerate}

\subsection*{Tuning the microcanonical demon}
\myindent
In the microcanonical method \emph{one} YM
configuration is reduced to a Polyakov loop configuration to start
the microcanonical run. Therefore the method is \emph{highly sensitive} to the
chosen starting configuration. Thus, the effect of choosing specific (well
thermalized) configurations with $\left. S_i\right\rvert_\text{config}$ in the
vicinity of $\ev{S_i}_{\text{YM}}$ is analyzed below.

For small $E_0^i$ thermalization problems of the demon arise due to the small
acceptance in the update procedure. If $E_0^i$ is too large the demon is able to
(and generically does) take away much energy
from the effective system. Thus, configurations within
the microcanonical ensemble become independent of the starting
configuration after the reduction step. These problems are circumvented in the
following way:
\begin{enumerate}
  \setlength{\itemsep}{0pt}
  \item Choose a large energy range $[-E_0^i,E_0^i]$ of the order $\ord(\abs{\ev{S^i}_\YM})$.
  \item Reduce the YM configuration to the Polyakov loop configuration $C_0$.
  \item For a few times ($10$ in our case) perform microcanonical
  simulations with $C_0$ as input configuration in every run. The demons' start
  energies are given by the expectation value $\ev{\Ed^i}$ in the preceding run. 
  \item The final run lasts for the same Monte-Carlo time as the preceding runs
  and is used to measure $\ev{\Ed^i}(\beta_i)$.
\end{enumerate}

Finally the contact between microscopic and microcanonical
system is based \emph{only on one configuration}. Further improvements
should be
possible by using a \emph{canonical demon} method with improved ``thermal contact''
to the microscopic system.

\section{Canonical demon method}
\myindent
In order to use the full statistics of the microscopic system we
apply the following algorithm \cite{Hasenbusch:1994zz}:
\begin{enumerate}
  \setlength{\itemsep}{0pt}
  \item Simulate the microscopic system according to $e^{-\Sw}$
  until thermalization.
  \item\label{it:simMC} Perform the reduction of the microscopic system to the effective system.
  \item Perform \emph{$N_\text{micro}$ microcanonical updates} of the joined
  system of effective model and demon energies. These updates do \emph{not}
  change the total energy $S^i+\Ed^i$.
  \item Freeze the demon system and update the microscopic fields up to a new
  independent configuration. After that proceed again with step \ref{it:simMC}.
\end{enumerate}
To deal with thermalization effects of the demons' energies we begin the
measurement after $N_\text{thermal}$ microscopic configurations with a suitably
chosen $N_\text{thermal}$.

\section{Observables for $\mathbf{SU(3)}$}
\myindent
We discuss the YM theory on a $\Ns^3\!\times\!\Nt$-lattice. The
Polyakov loop $\trP_\vcx$ is measured in terms of its lattice
average,\\[-2.3ex]
\begin{floatingfigure}[r]\begin{minipage}{0.45\linewidth}%
%
%
\begin{psfrags}%
\psfragscanon%
%
\psfrag{s03}[l][l]{\color[rgb]{0,0,0}\setlength{\tabcolsep}{0pt}\begin{tabular}{l}\scriptsize$\mathcal{F}$\end{tabular}}%
\psfrag{s04}[l][l]{\color[rgb]{0,0,0}\setlength{\tabcolsep}{0pt}\begin{tabular}{l}\scriptsize$\mathcal{F}$\end{tabular}}%
\psfrag{s05}[l][l]{\color[rgb]{0,0,0}\setlength{\tabcolsep}{0pt}\begin{tabular}{l}\scriptsize$\mathcal{F'}$\end{tabular}}%
\psfrag{s06}[l][l]{\color[rgb]{0,0,0}\setlength{\tabcolsep}{0pt}\begin{tabular}{l}\scriptsize$\mathcal{F''}$\end{tabular}}%
\psfrag{s07}[c][l]{\color[rgb]{0,0,0}\setlength{\tabcolsep}{0pt}\begin{tabular}{l}\scriptsize$\mathcal{F''}$\end{tabular}}%
\psfrag{s08}[Br][Bc]{\color[rgb]{0,0,0}\setlength{\tabcolsep}{0pt}\begin{tabular}{l}\scriptsize$\mathcal{F'}$\end{tabular}}%
\psfrag{s09}[t][t]{\color[rgb]{0,0,0}\setlength{\tabcolsep}{0pt}\begin{tabular}{c}\small $\Re
P$\end{tabular}}%
\psfrag{s10}[b][b][1][270]{\color[rgb]{0,0,0}\setlength{\tabcolsep}{0pt}\begin{tabular}{c}\small $\Im
P$\end{tabular}}
\psfrag{x01}[t][t]{\scriptsize$-2$}%
\psfrag{x02}[t][t]{\scriptsize$-1$}%
\psfrag{x03}[t][t]{\scriptsize$0$}%
\psfrag{x04}[t][t]{\scriptsize$1$}%
\psfrag{x05}[t][t]{\scriptsize$2$}%
\psfrag{x06}[t][t]{\scriptsize$3$}%
\psfrag{x07}[t][t]{\scriptsize$4$}%
%
\psfrag{v01}[r][r]{\scriptsize\phantom{$-2.5$}}%
\psfrag{v02}[r][r]{\scriptsize$-2$}%
\psfrag{v03}[r][r]{\scriptsize\phantom{$-1.5$}}%
\psfrag{v04}[r][r]{\scriptsize$-1$}%
\psfrag{v05}[r][r]{\scriptsize\phantom{$-0.5$}}%
\psfrag{v06}[r][r]{\scriptsize$0$}%
\psfrag{v07}[r][r]{\scriptsize\phantom{$0.5$}}%
\psfrag{v08}[r][r]{\scriptsize$1$}%
\psfrag{v09}[r][r]{\scriptsize\phantom{$1.5$}}%
\psfrag{v10}[r][r]{\scriptsize$2$}%
\psfrag{v11}[r][r]{\scriptsize\phantom{$2.5$}}%
%
\includegraphics[width=0.9\linewidth]{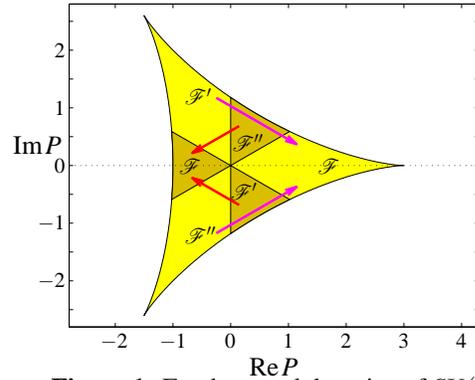}%
\end{psfrags}%
%
%
\end{minipage}\vspace*{0.5ex}%
\caption{\label{fig:rotatedObs}Fundamental domains of
$SU(3)$.}\end{floatingfigure}%
\noindent
\begin{minipage}{0.53\linewidth}\begin{equation}
P \equiv \frac{1}{V}\sum_\vcx \trP_\vcx,\quad V=\Ns^3.
\end{equation}\end{minipage}

Since we  deal
with phases where the traced
Polyakov loop is located halfway between the $SU(3)$ center
elements we project the value of the traced Polyakov loop onto
the nearest $\bbZ_3$-axis and define a \emph{rotated Polyakov loop} by (see Fig.~\ref{fig:rotatedObs})\\
\begin{minipage}{0.53\linewidth}
\begin{equation}
 \Prot =
 \begin{cases}
  \phantom{-\frac{1}{2}}\Re P  & \colon
  P\in \mathcal{F} \\
  -\frac{1}{2}\Re P + \frac{\sqrt{3}}{2}\Im P  & \colon
  P\in \mathcal{F'}  \\
  -\frac{1}{2}\Re P-\frac{\sqrt{3}}{2}\Im P  & \colon
  P\in \mathcal{F''}
  \end{cases} .
\end{equation}
\end{minipage}

\section{The $\mathbf{SU(3)}$ YM phase diagram with adjoint potential}
\myindent
An adjoint particle with mass
$M$ and spin $s$ at temperature $T$ leads to an effective potential 
\cite{Myers:2007vc}
\begin{equation}
\Delta V_\text{eff}=-\left[\frac{(2s+1)M^2T^2}{\pi^2} K_2(M/T) \chi_{11}(\trP)\right] = T\,
h\, \chi_{11}(\trP),\quad h<0.
\end{equation}
Nevertheless topological excitations allow for a positive $h$ and we therefore
study a lattice action 
\begin{equation}
\label{eq:su4action}
S = \beta \sum_{\square} \left( 1-  \frac{1}{N_C} \Re \, \tr
 \, U_{\square} \right)+ H\sum_\vcx \chi_{11}(\trP_\vcx)
\end{equation}
with standard $SU(3)$ Wilson action and adjoint potential
with unconstrained parameter $H$.

\begin{floatingfigure}
%
\begin{psfrags}%
\psfragscanon%
\psfrag{a000}[Br][Br]{\footnotesize $0.00$}%
\psfrag{a005}[Br][Br]{\footnotesize $0.05$}%
\psfrag{a00}[tr][tr]{\footnotesize $0.0$}%
\psfrag{a010}[Br][Br]{\footnotesize $0.10$}%
\psfrag{a015}[Br][Br]{\footnotesize $0.15$}%
\psfrag{a020}[Br][Br]{\footnotesize $0.20$}%
\psfrag{a025}[Br][Br]{\footnotesize $0.25$}%
\psfrag{a030}[Br][Br]{\footnotesize $0.30$}%
\psfrag{a05}[tr][tr]{\footnotesize $0.5$}%
\psfrag{a10}[tr][tr]{\footnotesize $1.0$}%
\psfrag{a15}[tr][tr]{\footnotesize $1.5$}%
\psfrag{a20}[tr][tr]{\footnotesize $2.0$}%
\psfrag{a25}[tr][tr]{\footnotesize $2.5$}%
\psfrag{a30}[tr][tr]{\footnotesize $3.0$}%
\psfrag{a50}[tr][tr]{\footnotesize $5.0$}%
\psfrag{a55}[tr][tr]{\footnotesize $5.5$}%
\psfrag{a60}[tr][tr]{\footnotesize $6.0$}%
\psfrag{a65}[tr][tr]{\footnotesize $6.5$}%
\psfrag{a70}[tr][tr]{\footnotesize $7.0$}%
\psfrag{m05}[tr][tr]{\footnotesize $-0.5$}%
\psfrag{x}[tr][tr]{\rule{0pt}{2ex}$\beta$}
\psfrag{y}[tr][tr]{$H$\rule{2em}{0pt}}
\psfrag{z}[tr][tr]{$\ev{\Prot}$\rule{1.2em}{0pt}}
\rule{0.05\columnwidth}{0pt}\includegraphics[width=0.5\columnwidth]{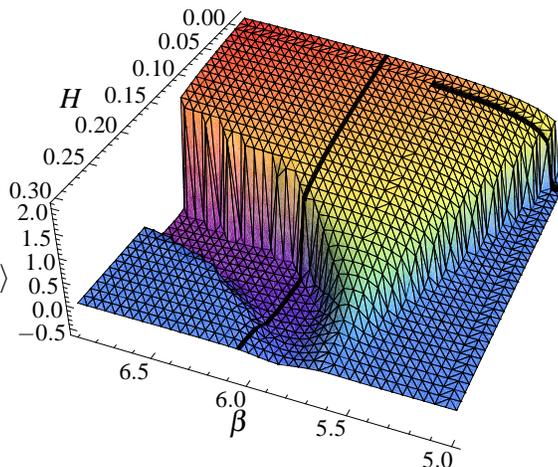}
\end{psfrags}

\caption{\label{fig:phasediagram}Phase diagram of $SU(3)$ YM with adjoint
potential according to Eq.~\eqref{eq:su4action}.}
\end{floatingfigure}
Simulations of this system near the
confinement-deconfinement phase transition on a $12^3\!\times\!2$ lattice
with varying $\beta$ and $H$ show the phase diagram
(Fig.~\ref{fig:phasediagram}) in terms of the rotated Polyakov loop $\Prot$. At
$H=0$ the well-known undirected and center-directed Polyakov loop structures
related  to (de)confinement appear. In the lower half-plane an additional
structure arises  where the Polyakov loop points into \emph{``anti-center''
direction}.\\ \indent
The additional potential term is already contained in the
effective model \eqref{eq:su3model}. We therefore not only analyze the
(de)confinement transition, but also look for a sensible  analysis of
the anti-center phase.

\subsection{The confinement-deconfinement transition}
\myindent
After simulating the (de)confinement phase transition (the upper black curve in
Fig.~\ref{fig:phasediagram}) we used IMC with the SD equations as well
as the (micro)canonical demon me\-thod for deriving couplings of the truncated
effective theory. The programming codes were
checked by simulating
effective theories with fixed couplings and \emph{reproducing them consistently} with
the SD equations and demon methods.

\begin{figure}[hb]
\begin{psfrags}%
\psfragscanon%
\psfrag{Prot}[cr][cr]{\small $\ev{\Prot}$}%
\psfrag{beta}{\small $\beta$}%
\psfrag{YM}[cr][cr]{\scriptsize Yang-Mills}%
\psfrag{SD}[cr][cr]{\scriptsize  Schw.-Dyson}%
\psfrag{microcan. demon}[cr][cr]{\scriptsize microc. demon}%
\psfrag{can. demon}[cr][cr]{\scriptsize can. demon}%
\psfrag{ 5}[cc][cc]{\scriptsize $5.00$}%
\psfrag{ 5.05}[cc][cc]{\scriptsize $5.05$}%
\psfrag{ 5.1}[cc][cc]{\scriptsize $5.10$}%
\psfrag{ 5.15}[cc][cc]{\scriptsize $5.15$}%
\psfrag{ 5.2}[cc][cc]{\scriptsize $5.20$}%
\psfrag{0.0}{\scriptsize $0.0$}%
\psfrag{0.2}{\scriptsize $0.2$}%
\psfrag{0.4}{\scriptsize $0.4$}%
\psfrag{0.6}{\scriptsize $0.6$}%
\psfrag{0.8}{\scriptsize $0.8$}%
\psfrag{1.0}{\scriptsize $1.0$}%
\psfrag{1.2}{\scriptsize $1.2$}%
\psfrag{1.4}{\scriptsize $1.4$}%
\includegraphics[width=0.49\columnwidth]{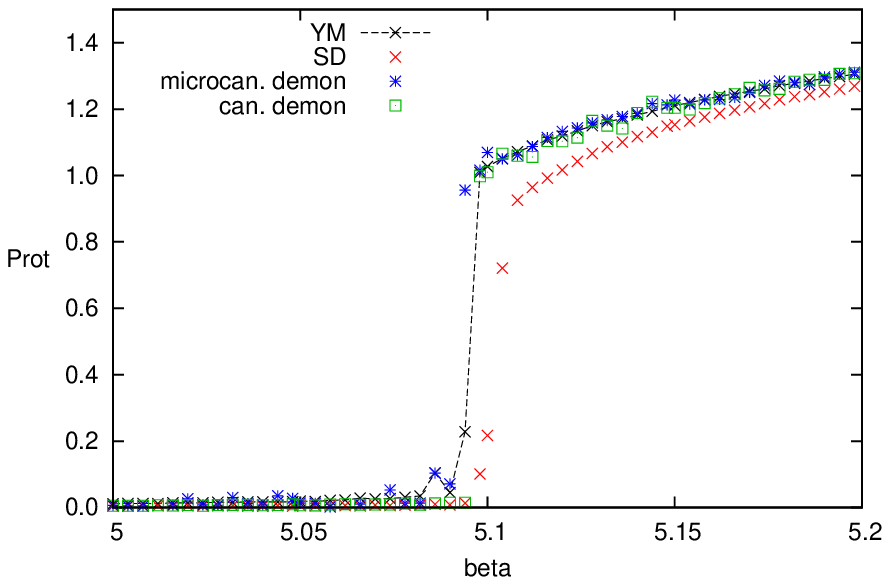}
\end{psfrags}
\hfill\begin{psfrags}%
\psfragscanon%
\psfrag{Prot}[cc][cr]{\small $\ev{\Prot}$}%
\psfrag{H}{\small $H$}%
\psfrag{YM}[cr][cr]{\scriptsize Yang-Mills}%
\psfrag{SD}[cr][cr]{\scriptsize  Schw.-Dyson}%
\psfrag{microcan. demon}[cr][cr]{\scriptsize microc. demon}%
\psfrag{can. demon}[cr][cr]{\scriptsize can. demon}%
\psfrag{ 0}[cc][cc]{\scriptsize $0.00$}%
\psfrag{ 0.05}[cc][cc]{\scriptsize $0.05$}%
\psfrag{ 0.1}[cc][cc]{\scriptsize $0.10$}%
\psfrag{ 0.15}[cc][cc]{\scriptsize $0.15$}%
\psfrag{ 0.2}[cc][cc]{\scriptsize $0.20$}%
\psfrag{ 0.25}[cc][cc]{\scriptsize $0.25$}%
\psfrag{ 0.3}[cc][cc]{\scriptsize $0.30$}%
\psfrag{-0.5}[cr][cr]{\scriptsize $-0.5$}%
\psfrag{0.0}[cr][cr]{\scriptsize $0.0$}%
\psfrag{0.5}[cr][cr]{\scriptsize $0.5$}%
\psfrag{1.0}[cr][cr]{\scriptsize $1.0$}%
\psfrag{1.5}[cr][cr]{\scriptsize $1.5$}%
\psfrag{2.0}[cr][cr]{\scriptsize $2.0$}%
\includegraphics[width=0.49\columnwidth]{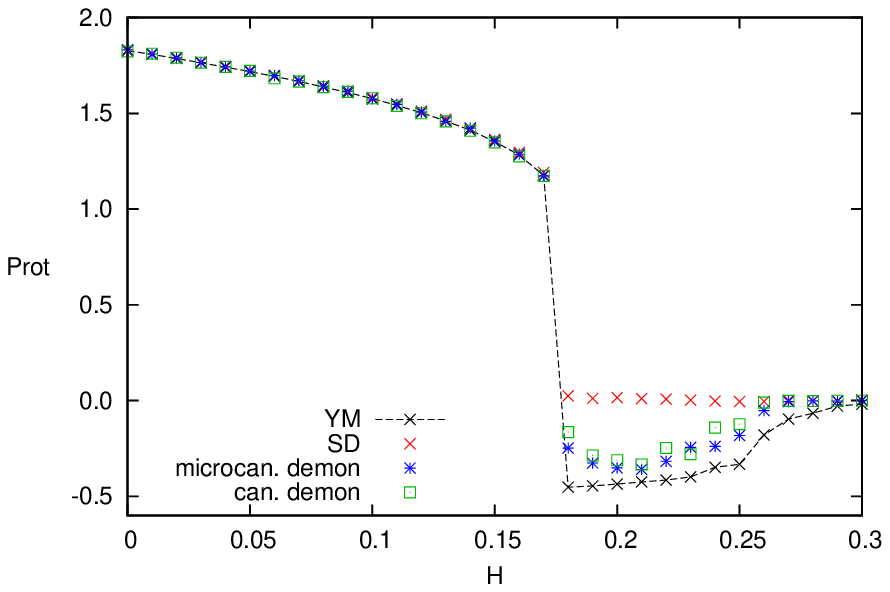}
\end{psfrags}
\caption{\label{fig:centerPhase}Expectation values of the full theory compared
to values produced with effective models after applying IMC methods.  \textsl{Left:}
(De)confinement
transition at $H=0$. \textsl{Right:} Anti-center phase at $\beta=6.05$.}
\end{figure}

The computed couplings corresponding to one YM coupling $\beta$ are
then used to simulate the associated effective theories with a Metropolis
algorithm. The resulting expectation values of the rotated Polyakov loop are
given in Fig.~\ref{fig:centerPhase} (left panel).
Here the SD equations fail
to reproduce the phase transition point whereas the
demon
methods \emph{reproduce
$\ev{\Prot}$ near the phase transition}
showing a better behavior than the SD method in the vicinity of the critical
coupling.

\subsection{The anti-center phase}
\myindent
At $\beta=6.05$ (vertical black curve in Fig.~\ref{fig:phasediagram})
we analyzed the anti-center phase and compared the resulting expectation values
of $\Prot$. 
Whereas the methods show a quantitative difference in rendering the
critical point near the (de)confinement transition the expectation values of
$\Prot$ indicate a \emph{qualitative difference} in the present case
(Fig.~\ref{fig:centerPhase}, right panel).
While the SD equations produce a smooth behavior with high accuracy deep
in the deconfined phase they fail to reproduce the anti-center phase
completely. In contrast demon methods are \emph{sensible to the full phase
structure}. This behavior in the anti-center phase was analyzed with
$750$ microcanonical runs at $\beta=6.05$, $H=0.2$ using different randomly chosen
starts. The resulting couplings are plotted in the three coupling phase diagram of the
effective theory
which shows a symmetric, center directed and anti-center
directed phase (see Fig.~\ref{fig:phaseBoundary}). Even far away from a phase
transition in the microscopic theory the
corresponding effective theory can be located in the vicinity of a phase
transition of the effective model.

\section{Using thermalized configurations for the
microcanonical demon}
\myindent
In the microcanonical demon method the starting
configuration can be chosen randomly from the full YM ensemble. In contrast we
can take configurations with $\left. S_i\right\rvert_\text{config}\approx
\ev{S_i}_{\text{YM}}$ (``well thermalized''). In the deconfined phase at
$\beta=6.05$, $H=0$ the projection of the three couplings to the
$\lambda_1$-$\lambda_2$ plane comparing the different starts is shown in
Fig.~\ref{fig:thermalization} (left
panel).
\begin{floatingfigure}[r]
\begin{psfrags}%
\psfragscanon%
\psfrag{lambda1}[cr][cr]{$\lambda_1$\rule{1em}{0pt}}%
\psfrag{lambda2}[cr][cr]{$\lambda_2$\rule{1.5ex}{0pt}}%
\psfrag{lambda3}[cr][cl]{$\lambda_3$\rule{0.7em}{0pt}}%
\psfrag{phase boundaries}[cr][cr]{\footnotesize phase boundaries}%
\psfrag{microcanonical couplings}[cr][cr]{\footnotesize microcan. couplings}%
\psfrag{microcan. demon}[cr][cr]{\footnotesize microcan. demon}%
\psfrag{can. demon}[cr][cr]{\footnotesize can. demon}%
\psfrag{symm}[cr][cr]{\small\textcolor{blue}{symmetric}}%
\psfrag{ferr}[cr][cr]{\small\textcolor{doran}{center}}%
\psfrag{antic}[cl][cl]{\small\textcolor{viol}{anti-center}}%
\psfrag{0.08}[cc][cr]{\footnotesize $0.08$\rule{2ex}{0pt}}%
\psfrag{0.07}[cc][cr]{\footnotesize $0.07$\rule{2ex}{0pt}}%
\psfrag{0.06}[cc][cr]{\footnotesize $0.06$\rule{2ex}{0pt}}%
\psfrag{0.05}[cc][cr]{\footnotesize $0.05$\rule{2ex}{0pt}}%
\psfrag{0.04}[cc][cr]{\footnotesize $0.04$\rule{2ex}{0pt}}%
\psfrag{0.03}[cc][cr]{\footnotesize $0.03$\rule{2ex}{0pt}}%
\psfrag{0.02}[cc][cb]{\footnotesize $0.02$\rule{0pt}{2ex}}%
\psfrag{0.00}[cc][cb]{}%
\psfrag{-0.02}[cc][cb]{\footnotesize $-0.02$\rule{0pt}{2ex}}%
\psfrag{-0.04}[cc][cb]{}%
\psfrag{-0.06}[cc][cb]{\footnotesize $-0.06$\rule{0pt}{2ex}}%
\psfrag{-0.08}[cc][cb]{}%
\psfrag{-0.10}[cc][cb]{\footnotesize $-0.10$\rule{0pt}{2ex}}%
\psfrag{-0.26}[cc][cr]{\footnotesize $-0.26$\rule{3ex}{0pt}}%
\psfrag{-0.28}[cc][cr]{\footnotesize $-0.28$\rule{3ex}{0pt}}%
\psfrag{-0.30}[cc][cr]{\footnotesize $-0.30$\rule{3ex}{0pt}}%
\psfrag{-0.32}[cc][cr]{\footnotesize $-0.32$\rule{3ex}{0pt}}%
\psfrag{-0.34}[cc][cr]{\footnotesize $-0.34$\rule{3ex}{0pt}}%
\psfrag{-0.27}[cc][cr]{}%
\psfrag{-0.29}[cc][cr]{}%
\psfrag{-0.31}[cc][cr]{}%
\psfrag{-0.33}[cc][cr]{}%
\psfrag{-0.35}[cc][cr]{}%
\includegraphics[width=0.5\columnwidth]{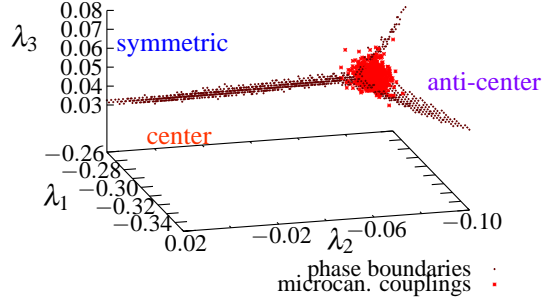}%
\end{psfrags}%
\caption{\label{fig:phaseBoundary}Phase boundaries of the three coupling
effective model and couplings derived from the microcanonical method at
$\beta=6.05$ and $H=0.2$.}
\end{floatingfigure}

The extent of the coupling distribution shrinks for the well thermalized
configurations compared to the randomly chosen configurations. Additionally
the couplings of the well thermalized configurations
correspond to
almost the same rotated Polyakov loop whereas the couplings derived from the
randomly chosen configurations show a much broader distribution of the
corresponding rotated Polya\-kov loops.

\section{Thermalization effects of the canonical demon}
\begin{figure}[h]
\begin{psfrags}%
\psfragscanon%
\psfrag{lambda1}[cc][cr]{$\lambda_1$}%
\psfrag{lambda2}[lc][cc]{$\lambda_2$}%
\psfrag{random start}[cr][cr]{\scriptsize random start}%
\psfrag{well thermalized}[cr][cr]{\scriptsize well thermalized}%
\psfrag{0.02}[cc][cc]{\footnotesize $0.02$\rule{1ex}{0pt}}%
\psfrag{0.01}[cc][cc]{}%
\psfrag{0.00}[cc][cc]{\footnotesize $0.00$\rule{1ex}{0pt}}%
\psfrag{-0.01}[cc][cc]{}%
\psfrag{-0.02}[cc][cc]{\footnotesize $-0.02$\rule{1ex}{0pt}}%
\psfrag{-0.03}[cc][cc]{}%
\psfrag{-0.04}[cc][cc]{\footnotesize $-0.04$\rule{1ex}{0pt}}%
\psfrag{-0.05}[cc][cc]{}%
\psfrag{-0.06}[cc][cc]{\footnotesize $-0.06$\rule{1ex}{0pt}}%
\psfrag{-0.07}[cc][cc]{}%
\psfrag{-0.08}[cc][cc]{\footnotesize $-0.08$\rule{1ex}{0pt}}%
\psfrag{-0.32}[cc][cc]{}%
\psfrag{-0.30}[cc][cc]{\footnotesize $-0.30$}%
\psfrag{-0.28}[cc][cc]{}%
\psfrag{-0.26}[cc][cc]{\footnotesize $-0.26$}%
\psfrag{-0.24}[cc][cc]{}%
\psfrag{-0.22}[cc][cc]{\footnotesize $-0.22$}%
\psfrag{-0.20}[cc][cc]{}%
\psfrag{-0.18}[cc][cc]{\footnotesize $-0.18$}%
\includegraphics[width=0.5\columnwidth]{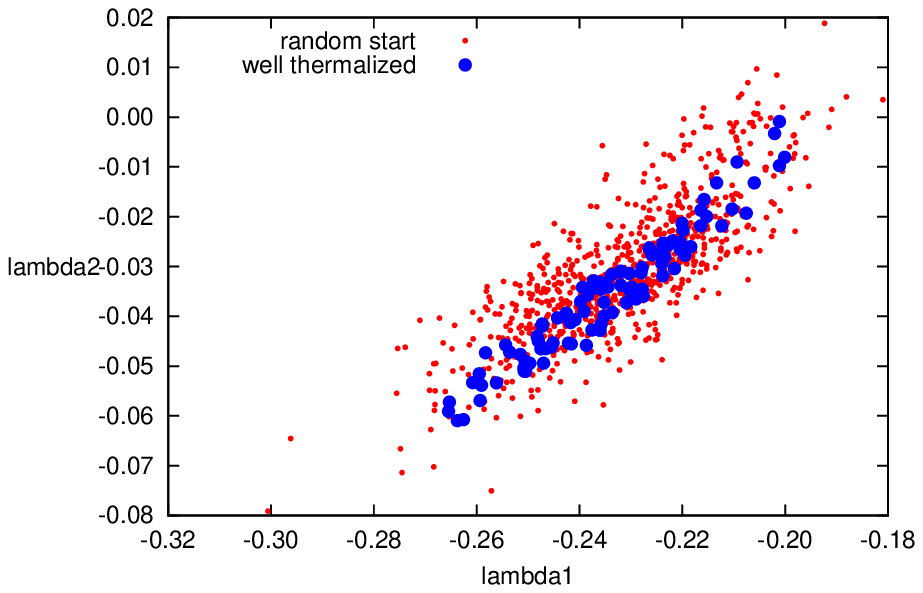}
\end{psfrags}
\hfill
\begin{psfrags}%
\psfragscanon%
\psfrag{lambda1}[cc][cr]{$\lambda_1$}%
\psfrag{lambda2}[lc][cc]{$\lambda_2$}%
\psfrag{bbbbb microcan.}[cr][cr]{\scriptsize microcan.\rule[-1ex]{0pt}{1em}}%
\psfrag{NM10}[cr][cr]{\scriptsize $N_\text{micro}=10$\rule[-0.6ex]{0pt}{1em}}%
\psfrag{NM100}[cr][cr]{\scriptsize $N_\text{micro}=10^2$}%
\psfrag{NM1000}[cr][cr]{\scriptsize $N_\text{micro}=10^3$}%
\psfrag{NM10000}[cr][cr]{\scriptsize
$N_\text{micro}=$\raisebox{-0.3ex}{$\,10^4$}\rule{0pt}{0.9em}}%
\psfrag{SD}[cr][cr]{\scriptsize Schw.-Dyson\rule{0pt}{1em}}%
\psfrag{0.02}[cc][cc]{\footnotesize $0.02$\rule{1ex}{0pt}}%
\psfrag{0.01}[cc][cc]{}%
\psfrag{0.00}[cc][cc]{\footnotesize $0.00$\rule{1ex}{0pt}}%
\psfrag{-0.01}[cc][cc]{}%
\psfrag{-0.02}[cc][cc]{\footnotesize $-0.02$\rule{1ex}{0pt}}%
\psfrag{-0.03}[cc][cc]{}%
\psfrag{-0.04}[cc][cc]{\footnotesize $-0.04$\rule{1ex}{0pt}}%
\psfrag{-0.05}[cc][cc]{}%
\psfrag{-0.06}[cc][cc]{\footnotesize $-0.06$\rule{1ex}{0pt}}%
\psfrag{-0.07}[cc][cc]{}%
\psfrag{-0.08}[cc][cc]{\footnotesize $-0.08$\rule{1ex}{0pt}}%
\psfrag{-0.32}[cc][cc]{}%
\psfrag{-0.30}[cc][cc]{\footnotesize $-0.30$}%
\psfrag{-0.28}[cc][cc]{}%
\psfrag{-0.26}[cc][cc]{\footnotesize $-0.26$}%
\psfrag{-0.24}[cc][cc]{}%
\psfrag{-0.22}[cc][cc]{\footnotesize $-0.22$}%
\psfrag{-0.20}[cc][cc]{}%
\psfrag{-0.18}[cc][cc]{\footnotesize $-0.18$}%
\includegraphics[width=0.5\columnwidth]{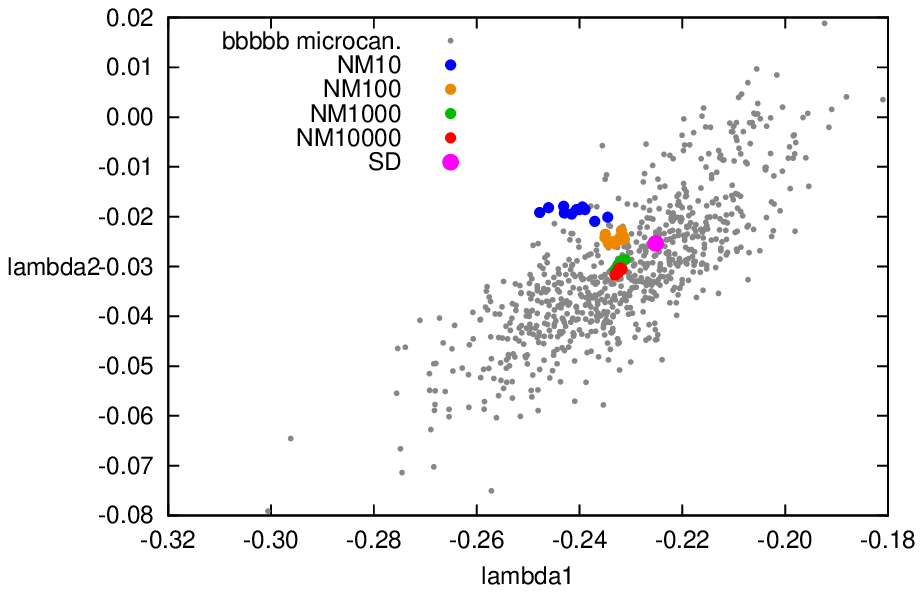}
\end{psfrags}
\caption{\label{fig:thermalization}\textsl{Left:} Couplings obtained via the
microcanonical demon method with randomly chosen and well thermalized
configurations.
\textsl{Right:} Thermalization effects due to different $N_\text{micro}$ of the
canonical demon method.}
\end{figure}

\myindent
One parameter of the canonical demon method is the number of
microcanonical sweeps per microscopic configuration $N_\text{micro}$. We
analyzed the distribution of the couplings in order to read off
any thermalization effects (Fig.~\ref{fig:thermalization}, right panel). The
microcanonical distribution serves as a reference.

Obviously the thermalization of the effective model has
to be taken into account. With a small $N_\text{micro}$ not enough time is spent
to thermalize the effective system completely. The measured couplings
refer to a non-equilibrium state of the effective model. Only in the large
$N_\text{micro}$ limit the couplings describe the thermal equilibrium of the
effective model. This behavior is given
by the fact that a configuration taken from a thermalized ensemble of the
microscopic (YM) system is \emph{not} necessarily a representative of
an equilibrium state of the effective theory \cite{Tomboulis:2007rn}.
Additionally the
couplings obtained via Schwinger-Dyson
equations do \emph{not correspond} to the ones computed with the canonical
demon method.

\begin{floatingfigure}[r]%
\begin{psfrags}%
\psfragscanon%
\psfrag{beta}[cr][cr]{$\beta$\rule{0pt}{1.5em}}%
\psfrag{absP}[cc][cc]{$\ev{\abs{P}}$\rule{0pt}{3em}}%
\psfrag{YM}[cr][cr]{\scriptsize Yang-Mills}%
\psfrag{can. demon}[cr][cr]{\scriptsize can. demon}%
\psfrag{0.0}[cc][cc]{\footnotesize $0.0$}%
\psfrag{0.5}[cc][cc]{\footnotesize $0.5$}%
\psfrag{1.0}[cc][cc]{\footnotesize $1.0$}%
\psfrag{1.5}[cc][cc]{\footnotesize $1.5$}%
\psfrag{2.0}[cc][cc]{\footnotesize $2.0$}%
\psfrag{9.60}[cc][cc]{\footnotesize $9.60$}%
\psfrag{9.65}[cc][cc]{\footnotesize $9.65$}%
\psfrag{9.70}[cc][cc]{\footnotesize $9.70$}%
\psfrag{9.75}[cc][cc]{\footnotesize $9.75$}%
\psfrag{9.80}[cc][cc]{\footnotesize $9.80$}%
\psfrag{9.85}[cc][cc]{\footnotesize $9.85$}%
\psfrag{9.90}[cc][cc]{\footnotesize $9.90$}%
\psfrag{9.95}[cc][cc]{\footnotesize $9.95$}%
\psfrag{10.00}[cc][cc]{\footnotesize $10.00$}%
\includegraphics[width=0.5\columnwidth]{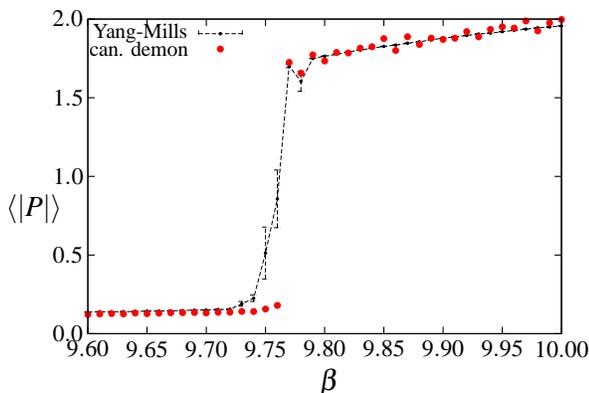}%
\end{psfrags}%
\caption{\label{fig:su4}Finite temperature phase transition of $SU(4)$ YM on a
$6^3\!\times\! 2$ lattice.}%
\end{floatingfigure}

\section{Outlook to $\mathbf{SU(4)}$ YM}
\myindent
For the finite temperature phase transition of the $SU(4)$ YM theory on a
$6^3\!\times\! 2$ lattice we applied the IMC method with the canonical demon
method. The effective model is a generalization of the three coupling model
for $SU(3)$ YM,\pagebreak
\begin{equation}
\begin{aligned}
\Seff &= \lambda_1
\sum_{\nN{\vcx\vcy}}(\chi_{100}(\trP_\vcx)\chi_{001}(\trP_\vcy)+\coco)
 + \lambda_2
 \sum_{\nN{\vcx\vcy}}(\chi_{010}(\trP_\vcx)\chi_{010}(\trP_\vcy)+\coco)\\ &\quad + \lambda_3
 \sum_{\nN{\vcx\vcy}}(\chi_{200}(\trP_\vcx)\chi_{002}(\trP_\vcy)+\coco) 
 + \lambda_4
\sum_{\nN{\vcx\vcy}}(\chi_{100}(\trP_\vcx)\chi_{001}(\trP_\vcy)+\coco)^2 .
\end{aligned}
\end{equation}
Even for $SU(4)$ the demon method is a robust way of obtaining couplings of
effective Polyakov loop models describing the phase transition (see
Fig.~\ref{fig:su4}).

\section{Conclusions}
\myindent
We studied and compared two ways of obtaining couplings for effective Polyakov
loop models which are related to $SU(N)$ YM theories. 
 The application of the inverse Monte-Carlo me\-thod with SD equations to the
$SU(3)$ YM case leads to \emph{stable results only far away} from the phase
transition. In the vicinity of the phase transition demon methods lead to a much better
sampling of expectation values of the Polya\-kov loop. We tried to reproduce the
anti-center phase
of a model with standard Wilson action and adjoint Polya\-kov loop potential  by SD and demon methods.
The SD method fails to reco\-ver the
phase structure while demon methods are favorable even in this case.
The demon method can be \emph{generalized} straightforwardly to the case of
$SU(4)$ YM leading to robust results in the vicinity of the finite temperature
phase transition.
Combining our experiences with both methods SD equations are
less efficient than demon methods near
first order phase transitions although SD equations have proven to be very
useful for the second order transition in $SU(2)$ YM
\cite{Heinzl:2005xv}.

When using demon methods much care has to be taken of
different thermalization effects. Firstly with microcanonical demons the way of
choosing microscopic configurations influences the derived couplings. Secondly
when using the canonical demon method thermalization effects (which
cannot be cured as discussed in \cite{Tomboulis:2007rn}) must be taken into
account.

\acknowledgments
\myindent
We thank A.~Velytsky for an interesting discussion at the conference.
TK acknowledges support by the Konrad-Ade\-nauer-Stiftung
e.V. and CW by
the Studienstiftung des deutschen Volkes.
This work has been supported by the DFG grant Wi 777/8-2.

\newlength{\bibskip}
\setlength{\bibskip}{-0.55ex}

\end{document}